\documentclass[twoside,11pt]{7ssmmp} 

\usepackage{amsfonts}
\usepackage{amssymb}
\usepackage{fancyhdr}    
\usepackage{graphicx}
\usepackage{rotating}
\usepackage{amsxtra}
\usepackage{latexsym}
\usepackage{amsmath}
\usepackage{bm}

\newcommand{\be}{\begin{equation}}
\newcommand{\ee}{\end{equation}}
\newcommand{\bea}{\begin{eqnarray}}
\newcommand{\eea}{\end{eqnarray}}
\newcommand{\beaa}{\begin{eqnarray*}}
\newcommand{\eeaa}{\end{eqnarray*}}

\newcommand{\nn}{\nonumber \\}
\newcommand{\e}{\mathrm{e}}

\newcommand{\Eqn}[1]{&\hspace{-0.2em}#1\hspace{-0.2em}&}

\def\be{\begin{equation}}
\def\ee{\end{equation}}
\def\bea{\begin{eqnarray}}
\def\eea{\end{eqnarray}}

\def\nn{\nonumber \\}
\def\e{\mathrm{e}}

\begin{document}
 \baselineskip=11pt

\title{Modified gravity: walk through accelerating cosmology
\hspace{.25mm}\thanks{\,
Work supported in part by 
Global COE Program of Nagoya University (G07)
provided by the Ministry of Education, Culture, Sports, Science \&
Technology and by the JSPS Grant-in-Aid for Scientific Research (S) \# 22224003
and (C) \# 23540296 (S.N.);\ 
and 
MINECO (Spain), FIS2010-15640 and
AGAUR (Generalitat de Ca\-ta\-lu\-nya), contract 2009SGR-345
(S.D.O.). 
This paper is based on the lecture given by S.D. Odintsov at 
the 7th Mathematical Physics Meeting: Summer School and Conference on Modern 
Mathematical Physics, Belgrade, 9-19.09.2012.
}
}
\author{\bf{Kazuharu Bamba}\hspace{.25mm}\thanks{\,e-mail address: 
bamba@kmi.nagoya-u.ac.jp}
\\ \normalsize{Kobayashi-Maskawa Institute for the Origin of Particles and the
Universe,} \\ \normalsize{Nagoya University, Nagoya 464-8602, Japan} 
\vspace{2mm} \\ 
\bf{Shin'ichi Nojiri}\hspace{.25mm}\thanks{\,e-mail
address: nojiri@phys.nagoya-u.ac.jp}
\\ \normalsize{Department of Physics,} \\ \normalsize{Nagoya University, Nagoya 464-8602, Japan} \vspace{2mm} \\ 
\normalsize{Kobayashi-Maskawa Institute for the Origin of Particles and the
Universe,} \\ \normalsize{Nagoya University, Nagoya 464-8602, Japan} 
\vspace{2mm} \\ 
\bf{Sergei D. Odintsov}\hspace{.25mm}\thanks{\,e-mail
address: odintsov@ieec.uab.es}
\\ \normalsize{Instituci\`{o} Catalana de Recerca i Estudis Avan\c{c}ats (ICREA),} \\ \normalsize{Barcelona, Spain} 
\vspace{2mm} \\ 
\normalsize{Institut de Ciencies de l'Espai (CSIC-IEEC),} \\ \normalsize{
Campus UAB, Facultat de Ciencies,} \\ \normalsize{Torre C5-Par-2a pl, E-08193 Bellaterra
(Barcelona), Spain} 
\vspace{2mm} \\ 
\normalsize{Tomsk State Pedagogical University, Tomsk, Russia} 
\vspace{2mm} \\ 
\normalsize{Eurasian National University, Astana 010008, Kazakhstan} 
}

\date{}

\maketitle

\begin{abstract}
We review the accelerating (mainly, dark energy) cosmologies in modified 
gravity. Special attention is paid to cosmologies leading to 
finite-time future singularities in $F(R)$, $F(G)$ and $\mathcal{F}(R,G)$ 
modified gravities. The removal of the finite-time future singularities via 
addition of $R^2$-term which simultaneously unifies the early-time inflation 
with late-time acceleration is also briefly mentioned. Accelerating cosmology 
including the scenario unifying inflation with dark energy is considered in 
$F(R)$ gravity with Lagrange multipliers. 
In addition, we examine domain wall solutions in $F(R)$ gravity. 
Furthermore, covariant higher derivative gravity with scalar 
projectors is explored. 
\end{abstract}

\vspace{5mm}
PACS \, numbers: \,\, {04.50.Kd, 95.36.+x, 98.80.-k}
\\


\vspace{-7mm}

\section{Introduction}

It is observationally implied that the current expansion of the universe 
is accelerating. 
Provided that the universe is homogeneous, 
two representative approaches to account for the current 
cosmic acceleration exist. 
The first is to assume the existence of the so-called dark energy whose 
pressure is negative (for a recent review, see, 
e.g.,~\cite{Bamba:2012cp}).  
The second is to consider that a gravitational theory would be modified 
at the large distance scale. The simplest theory is $F(R)$ gravity 
(for reviews, see, for example,~\cite{Review-Nojiri-Odintsov-CF}). 

In this paper, we examine the accelerating (dark energy) solutions of modified 
gravity which may produce future singularities. 
We concentrate on reviewing the results in Refs.~\cite{Bamba:2009uf, Capozziello:2010uv, Bamba:2011nm, Kluson:2011rs} 
on theoretical aspects of modified gravity theories with presenting 
dark energy components. 
In particular, we study 
the finite-time future singularities in $F(R)$, $F(G)$ and 
$\mathcal{F}(R,G)$ gravity theories~\cite{Bamba:2009uf, Cognola:2007vq, DeFelice:2010hg}, where $R$ is the Ricci scalar, 
$G \equiv R^{2}-4R_{\mu\nu}R^{\mu\nu}+R_{\mu\nu\rho\sigma}R^{\mu\nu\rho\sigma}$
with $R_{\mu\nu}$ and $R_{\mu\nu\rho\sigma}$ the Ricci tensor and
the Riemann tensor, respectively, is the Gauss-Bonnet invariant, 
and $\mathcal{F}(R,G)$ is an arbitrary function of $R$ and $G$. 
This is a generalized gravity theory including both 
$F(R)$ and $F(G)$ gravity theories. 
We also discuss the removal of the finite-time future singularities in $F(R)$ 
gravity via addition of $R^2$-term which simultaneously leads to the 
unification of early-time inflation with late-time 
acceleration~\cite{Nojiri:2003ft}. 
In the frameworks of $F(G)$ or $\mathcal{F}(R,G)$ theory, 
the corresponding term may be different, of course~\cite{Bamba:2009uf}. 
We note that as related studies, 
the finite-time future singularities~\cite{Bamba:2008ut, Dev:2008rx, Bamba:2009vq} and the realization of the phantom phase including 
the crossing of the phantom divide~\cite{PC} have also been examined. 
Furthermore, the features of the finite-time future singularities in 
non-local gravity~\cite{Bamba:2012ky}, 
modified teleparallel gravity~\cite{Bamba:2012vg} 
and its extended analysis in loop quantum cosmology (LQC)~\cite{Bamba:2012ka} 
has recently been investigated. 
In addition, dark energy in the context of 
$F(R)$ gravity with Lagrange multipliers~\cite{Capozziello:2010uv} is 
considered. 
We also present domain wall solutions in $F(R)$ gravity~\cite{Bamba:2011nm}. 
Moreover, covariant higher derivative gravity with scalar 
projectors~\cite{Kluson:2011rs} is explained. 
We use units of $k_\mathrm{B} = c = \hbar = 1$ and denote the
gravitational constant $8 \pi G_\mathrm{N}$ by
${\kappa}^2 \equiv 8\pi/{M_{\mathrm{Pl}}}^2 = 1$
with the Planck mass of $M_{\mathrm{Pl}} = G_\mathrm{N}^{-1/2} = 1.2 \times
10^{19}$GeV.

The paper is organized as follows.
In Section 2, we explore accelerating cosmologies leading to the finite-time 
future singularities in $F(R)$, $F(G)$ and 
$\mathcal{F}(R,G)$ gravity theories. 
In Section 3, we study dark energy in the framework of 
$F(R)$ gravity with Lagrange multipliers. 
In Section 4, we examine domain wall solutions in $F(R)$ gravity. 
In Section 5, we investigate covariant higher derivative gravity with scalar 
projectors. 
Finally, conclusions are presented in Section 6.

\section{Finite-time future singularities in 
$F(R)$, $F(G)$ and $\mathcal{F}(R,G)$ gravity theories}

\subsection{$\mathcal{F}(R,G)$ gravity}

The action of $\mathcal{F}(R,G)$ gravity is 
$
S = \int d^4 x \sqrt{-g} \left[ \mathcal{F}(R,G)/\left(2\kappa^2\right)
+{\mathcal{L}}_{\mathrm{M}} \right]
$, where $g$ is the determinant of the metric tensor $g_{\mu\nu}$
and ${\mathcal{L}}_{\mathrm{M}}$ is the matter Lagrangian. 
This is a generic theory including both $F(R)$ and $F(G)$ gravities. 
We take the flat Friedmann-Lema\^{i}tre-Robertson-Walker (FLRW) metric 
%
$
ds^2 = - dt^2 + a^2(t) \sum_{i=1,2,3}\left(dx^i\right)^2
$. 
%
The Hubble parameter is given by 
$H=\dot{a}/a$, where 
the dot denotes the time derivative of $\partial/\partial t$. 
In the FLRW background, with the gravitational field equations 
we find that the effective (i.e., total) energy density and pressure of 
the universe read 
$\rho_{\mathrm{eff}}=3\kappa^{-2}H^{2}$ 
and 
$
P_{\mathrm{eff}}=-\kappa^{-2} \left( 2\dot H+3H^{2} \right)
$, respectively. 
For the action in Eq.~(\ref{eq:1}), we obtain 
\begin{eqnarray}
&& 
\hspace{-12mm}
\rho_{\mathrm{eff}} \equiv 
\frac{1}{{\mathcal{F}}_{,R}} \left\{ \rho_{\mathrm{M}} +
\frac{1}{2\kappa^{2}}
\left[ \left( {\mathcal{F}}_{,R}R-\mathcal{F} \right)
-6H{\dot{\mathcal{F}}}_{,R}
+G{\mathcal{F}}_{,G}-24H^3{\dot{\mathcal{F}}}_{,G}
\right] \right\}\,,
\label{eq:1} \\
&& 
\hspace{-12mm}
P_{\mathrm{eff}} \equiv 
\frac{1}{{\mathcal{F}}_{,R}} \left\{ P_{\mathrm{M}} +
\frac{1}{2\kappa^{2}} \left[
-\left( {\mathcal{F}}_{,R}R-\mathcal{F} \right)
+4H{\dot{\mathcal{F}}}_{,R}+2{\ddot{\mathcal{F}}}_{,R}
-G{\mathcal{F}}_{,G}
\right. 
\right. \nonumber \\ 
&& \left. \left. 
{}
+16H\left(\dot{H} +H^2 \right){\dot{\mathcal{F}}}_{,G}
+8H^2 {\ddot{\mathcal{F}}}_{,G}
\right]
\right\}\,,
\label{eq:2}
\end{eqnarray}
where 
$
{\mathcal{F}}_{,R} \equiv 
\partial \mathcal{F}(R,G)\partial R 
$ 
and 
$
{\mathcal{F}}_{,G} \equiv 
\partial \mathcal{F}(R,G)\partial G
$, 
and $\rho_{\mathrm{M}}$ and $P_{\mathrm{M}}$ are the 
energy density and pressure of matter (which has been 
assumed to be a perfect fluid).

\subsection{Finite-time future singularities}

Provided that the Hubble parameter is written as 
\begin{equation}
H=\frac{h_{\mathrm{s}}}{(t_{\mathrm{s}}-t)^{\beta}} +H_{\mathrm{s}}\,,
\label{eq:3}
\end{equation}
where $h_{\mathrm{s}} (>0)$, $t_{\mathrm{s}} (>0)$, $H_{\mathrm{s}} (\geq 0)$, 
and $\beta (\neq 0)$ are constants, 
$t_{\mathrm{s}}$ is the time when a finite-time future singularity occurs, 
and $0 < t < t_{\mathrm{s}}$. 
In what follows, we consider the case of $H_{\mathrm{s}} =0$. 
We note that 
even if $\beta<0$ and $\beta$ is a non-integer value, 
in the limit $t \to t_{\mathrm{s}}$ 
some derivative of $H$ diverges 
and hence the scalar curvature becomes infinity~\cite{Bamba:2008ut}. 
Moreover, since the case of $\beta=0$ leads to a de Sitter space, 
we suppose $\beta \neq 0$.

The finite-time future singularities are classified into four types~\cite{Nojiri:2005sx}. 
%
Type I (``Big Rip''~\cite{Caldwell:2003vq}):\ 
In the limit $t\to t_{\mathrm{s}}$, 
$a \to \infty$,
$\rho_{\mathrm{eff}} \to \infty$ and
$\left|P_{\mathrm{eff}}\right| \to \infty$. 
The case that 
$\rho_\mathrm{{eff}}$ and $P_{\mathrm{eff}}$ are finite values 
at $t = t_{\mathrm{s}}$~\cite{Shtanov:2002ek} 
is included. 
This happens for $\beta=1$ and $\beta>1$. 
In this paper, we regard the singularities for $\beta=1$ as 
``Big Rip'' and those for $\beta>1$ as ``Type I''. 
%
(ii) 
Type II (``sudden''~\cite{sudden}):\ 
In the limit $t\to t_{\mathrm{s}}$, 
$a \to a_{\mathrm{s}}$, 
$\rho_{\mathrm{eff}} \to \rho_{\mathrm{s}}$ and 
$\left|P_{\mathrm{eff}}\right| \to \infty$. 
This occurs for $-1<\beta<0$.
%
(iii)
Type III:\ 
In the limit $t\to t_{\mathrm{s}}$, 
$a \to a_{\mathrm{s}}$, 
$\rho_{\mathrm{eff}} \to \infty$ and
$\left|P_{\mathrm{eff}}\right| \to \infty$. 
This appears for $0<\beta<1$.
%
(iv) 
Type IV:\ 
In the limit $t\to t_{\mathrm{s}}$, 
$a \to a_{\mathrm{s}}$, 
$\rho_{\mathrm{eff}} \to 0$, 
$\left|P_{\mathrm{eff}}\right| \to 0$, 
and higher derivatives of $H$ diverge. 
The case that $\rho_{\mathrm{eff}}$ and/or $\left|P_{\mathrm{eff}}\right|$ 
become finite values at $t = t_{\mathrm{s}}$ is also included. 
This is realized if $\beta<-1$ but $\beta$ is not any integer number. 
%
Here, $a_{\mathrm{s}} (\neq 0)$ and $\rho_{\mathrm{s}}$ 
are constants. 
%

\subsection{$F(R)$ gravity with finite-time future singularities}

By taking $\mathcal{F}(R,G) = F(R)$, 
the action in Eq.~(\ref{eq:1}) becomes that of $F(R)$ gravity. 
With the method to reconstruct modified gravity~\cite{Bamba:2008ut, 
Reconstruction, Nojiri:2006gh}, for the Hubble parameter to be represented in 
Eq.~(\ref{eq:3}) we explore $F(R)$ gravity models in which 
finite-time future singularities can appear. 
By introducing two proper functions $P(\phi)$ and $Q(\phi)$ of a scalar field 
$\phi$, which we regard as the cosmic time $t$, 
we rewrite the term $\mathcal{F}(R,G) = P(t)R+Q(t)$ 
in the action in Eq.~({\ref{eq:1}}). 
In this case, by varying the action with respect to $t$ we acquire
$
\left(dP(t)/dt\right)R + dQ(t)/dt=0
$. In principle, by solving this equation we have the relation $t=t(R)$. 
If we substitute it into the above form of $\mathcal{F}(R,G) = P(t)R+Q(t)$, 
we find $F(R)=P(t=t(R))R+Q(t=t(R))$ and hence the original action 
is found again. 
We express the scale factor as
$
a(t)= \bar{a} \exp \left( \bar{g}(t) \right)
$ 
with $\bar{a}$ a constant and $\bar{g}(t)$ a proper function. 
Here, we neglect the contribution from matter 
because when the finite-time future singularities appears, 
the energy density of dark energy components are completely dominant over 
that of matter. In this case, the gravitational field equations yield 
\begin{equation}
\ddot{P}(t) - \dot{\bar{g}}(t) \dot{P}(t) + 2 \ddot{\bar{g}}(t) P(t) = 0\,, 
\quad 
Q(t) = -6 \left[ \left( \dot{\bar{g}}(t) \right)^2 P(t) + \dot{\bar{g}}(t) \dot{P}(t)
\right]\,.
\label{eq:4}
\end{equation}
Accordingly, if we find the solutions $P(t)$ and $Q(t)$ of these 
equations, by plugging those into $F(R)=P(t)R+Q(t)$ with $t=t(R)$ 
we obtain the concrete form of $F(R)$. 
We acquire the followings consequences. 

\noindent
(a) 
$\beta=1$ [Big Rip]:\ 
For 
$h_{\mathrm{s}} > 5 + 2\sqrt{6}$ or $h_{\mathrm{s}} < 5 - 2\sqrt{6}$, 
$F(R) \propto R^{q}$ with $q \equiv \left(1/4\right) \left(3+h_{\mathrm{s}} 
+\sqrt{h_{\mathrm{s}}^2 - 10h_{\mathrm{s}} +1} \right)$, 
whereas if $5 - 2\sqrt{6} < h_{\mathrm{s}} < 5 + 2\sqrt{6}$, 
$
F(R) \propto R^{\left(h_{\mathrm{s}} + 1\right)/4} \times 
\left( \mbox{Oscillating part} \right)
$. 

\noindent
(b) 
$\beta>1$ [Type I]:\ 
$F(R) \propto \exp \left\{
\left(h_{\mathrm{s}}/\left[2\left(\beta - 1\right)\right)\right]
\left(\frac{R}{12h_{\mathrm{s}}}\right)^{\left(\beta - 1\right)/\left(2\beta\right)} \right\}
R^{-1/4}\times \left( \mbox{Oscillating part} \right)
$.

\noindent
(c) 
$0 < \beta < 1$ [Type III]:\ 
$
F(R) \sim \exp \left[
\frac{h_{\mathrm{s}}}{2\left(\beta-1\right)}
\left( - 6\beta h_{\mathrm{s}} R \right)^{(\beta - 1)/(\beta + 1)} \right] R^{7/8}
$.

\noindent
(d) 
$\beta < 0$ [Type II ($-1<\beta<0$) and 
Type IV ($\beta<-1$ but $\beta$ is not an integer)]:\ 
$
F(R) \sim 
\left( -6h_{\mathrm{s}} \beta R \right)^{\left(\beta^2 + 2\beta + 9\right)/\left[8\left(\beta +1\right)\right]}
\exp \left[ 
\frac{h_{\mathrm{s}}}{2\left(\beta-1\right)} 
\left( -6h_{\mathrm{s}} \beta R \right)^{\left(\beta-1\right)/\left(\beta + 1\right)} 
\right]
$. 
Here, ``$\sim$'' means the asymptotic behavior in the limit 
$t \to t_{\mathrm{s}}$. 

It is remarkable that adding $R^2$-term to such a theory, 
one removes future singularity. 
(Note that $R^2$ gravity was proposed as inflationary model in 
Ref.~\cite{Starobinsky:1980te} (celebrated Starobinsky inflation) and 
was used for the first unified inflation-dark energy 
modified gravity proposed in Ref.~\cite{Nojiri:2003ft}). 
Hence, we not only 
remove singularities by adding $R^2$ term but also unify the dark energy era 
with inflation in such a way 
(for realistic models of such unification, see~\cite{Review-Nojiri-Odintsov-CF}). 
Several viable $F(R)$ gravity models which unify inflation with dark energy 
and do not contain the finite-time future singularities are listed 
below~\cite{Review-Nojiri-Odintsov-CF}: 
%
\begin{eqnarray}
&& 
\hspace{-10mm}
F(R) = R + \frac{b_1 R^{2l} - b_2 R^l}{1 + b_3 R^l} + b_4 R^2\,, 
\label{eq:Add-2.3-01} \\
%
%
&&
\hspace{-10mm}
F(R)=R-2\Lambda\left[1-\mathrm{e}^{-R/\left(b_5 \Lambda\right)}\right]
-\Lambda_\mathrm{i}\left[1-\mathrm{e}^{-\left(R/R_\mathrm{i}\right)^{\varsigma}}\right]
+b_6 \tilde R_\mathrm{i}^{-\left(\tau-1\right)} R^{\tau}\,, 
\label{eq:Add-2.3-02}
\end{eqnarray}
with $b_j$ ($j=1, \dotsc, 4$), $b_5 (>0)$, $b_6 (>0)$ and $l$ constants. 
In Eq.~(\ref{eq:Add-2.3-02}), 
$\tau (>1)$ is a natural number, $\varsigma$ and $\tilde{R}_\mathrm{i}$ are constants, 
$R_\mathrm{i}$ and $\Lambda_\mathrm{i}$ are 
transition curvature and expected cosmological constant 
at the inflationary stage, respectively~\cite{Elizalde:2010ts}.

\subsection{$F(G)$ gravity with finite-time future singularities}

With the same method as in $F(R)$ gravity in Section 2.3, 
it is possible to execute the reconstruction of 
$F(G)$ gravity models in which the finite-time future singularities 
occur. 
The action of $F(G)$ gravity~\cite{Nojiri:2005jg} 
is described by Eq.~(\ref{eq:1}) with $\mathcal{F}(R,G) = R+F(G)$.
In this case, the gravitational field equations yield 
\begin{equation}
2 \frac{d}{d t} \left( \dot{\bar{g}}^2(t)  \dot{P}(t) \right) -2
\dot{\bar{g}}^{3}(t) \dot{P}(t) + \ddot{\bar{g}}(t)=0\,, 
\quad 
Q(t)= -24 \dot{\bar{g}}^{3}(t) \dot{P}(t) - 6\dot{\bar{g}}^2(t)\,.
\label{eq:5}
\end{equation}
The results are as follows~\cite{Bamba:2009uf}. 

\noindent
(a) $\beta=1$ [Big Rip]:\ 
For $h_{\mathrm{s}} \neq 1$, 
$
F(G)=\left\{\sqrt{6h_{\mathrm{s}}^{3}(1+h_{\mathrm{s}})}/\left[h_{\mathrm{s}} (1-h_{\mathrm{s}})\right]\right\}\sqrt{G}
+c_{1}G^{\left(h_{\mathrm{s}} +1\right)/4}+c_{2}G
$ 
with $c_{1}$ and $c_{1}$ constants. 
If $h_{\mathrm{s}} = 1$, 
$
F(G)= \frac{\sqrt{3}}{2} 
\sqrt{G} \ln \left( \gamma G \right)
$ with $\gamma (>0)$ a positive constant. 

\noindent
(b) $\beta>1$ [Type I]:\ 
$F(G) = -\sqrt{6} \sqrt{G}$.

\noindent
(c) 
$0<\beta<1$ [Type III], $-1/3<\beta<0$ [Type II], $-1<\beta<-1/3$ [Type II] 
and $\beta<-1$ (but $\beta$ is not integer) [Type IV]:\ 
$
F(G)=6h_{\mathrm{s}}^{2}(3\beta+1)(\beta+1)^{-1}
\left[|G|/\left(24h_{\mathrm{s}}^{3}|\beta|\right)\right]^{2\beta/(3\beta+1)}
$. 

\noindent
(d) 
$\beta=-1/3$ [Type II] (this is a special value in this case):\ 
$
F(G) \simeq \left[1/\left(4 \sqrt{6}h_{\mathrm{s}}^{3}\right)\right]G\left(G+8h_{\mathrm{s}}^{3}\right)^{1/2}
+\left(2/\sqrt{6}\right)\left(G+8h_{\mathrm{s}}^{3}\right)^{1/2}
$. 
We remark that the finite-time future singularities 
appearing in the limit $G\rightarrow \pm\infty$ 
can be removed by the additional term $d_1 G^{\varrho}$, where $d_1 (\neq 0)$ 
is a constant, and $\varrho > 1/2$ and $\varrho \neq 1$. 
Furthermore, the finite-time future singularities emerging in the limit 
$G\rightarrow 0^{-}$ can be cured by adding 
the term $d_1 G^{\varrho}$, where $\varrho (\leq 0)$ 
is an integer~\cite{Bamba:2009uf}.

\subsection{$\mathcal{F}(R,G)$ gravity with finite-time future singularities}

Using the similar procedure in Section 2.3, 
we reconstruct the form of $\mathcal{F}(R,G)$ leading to 
the finite-time future singularities. 
With proper functions $P(\phi)$, $Z(\phi)$ and $Q(\phi)$ of a
scalar field $\phi$, which we indentify with $t$, 
we represent the term $\mathcal{F}(R,G)$ in the action in Eq.~({\ref{eq:1}}) 
as $P(t)R+Z(t)G+Q(t)$. 
Varying this action with respect to $t$, we find 
$\left( dP(t)/dt \right)R + \left( dZ(t)/dt \right)G + dQ(t)/dt = 0$. 
By solving this equation, we obtain $t=t(R,G)$. 
Combining this and the above representation $P(t)R+Z(t)G+Q(t)$, 
we acquire $\mathcal{F}(R,G) = P(t)R+Z(t)G+Q(t)$. 
It follows from the gravitational field equations, the conservation law, 
$a(t)= \bar{a} \exp \left( \bar{g}(t) \right)$, and 
$H(t)=\dot{\bar{g}}(t)$ 
that 
%
\begin{eqnarray}
&&
\frac{d^2 P(t)}{dt^2} + 4 \dot{\bar{g}}^{2}(t) \frac{d^2 Z(t)}{dt^2} 
-\dot{\bar{g}}(t) \frac{d P(t)}{dt} 
\nonumber \\ 
&&
{}+ 4\left(2 \dot{\bar{g}} \ddot{\bar{g}} - 
\dot{\bar{g}}^{3}(t)\right) \frac{d Z(t)}{dt} + 2\ddot{\bar{g}}(t)P(t) 
=0\,, 
\label{eq:6} \\
&&
Q(t)=-6 \left( 4\dot{\bar{g}}^{3}(t) \frac{d Z(t)}{dt} -\dot{\bar{g}}^{2}(t)P(t)-\dot{\bar{g}}(t) \frac{d P(t)}{dt} \right)\,.
\label{eq:7}
\end{eqnarray}
For $P(t) \neq 0$, $\mathcal{F}(R,G)$ can be described as 
$\mathcal{F}(R,G)=R \tilde{g}(R,G) + \tilde{f}(R,G)$ with 
$\tilde{g} (R,G) (\neq 0)$ and $\tilde{f} (R,G)$ 
generic functions of $R$ and $G$. 
We show the results. 

\noindent
(a) $\beta=1$ [Big Rip]:\ 
For 
$0< h_{\mathrm{s}}< 5-2\sqrt{6}$ or $h_{\mathrm{s}} > 2+\sqrt{6}$, 
we find 
$\mathcal{F}(R,G)=\alpha_{1}R^{q_{+}}+\alpha_{2}R^{q_{-}}
+\delta G^{\left(h_{\mathrm{s}}+1\right)/4}
$ 
with 
$q_{\pm} \equiv \left(1/4\right) \left(3+h_{\mathrm{s}} 
\pm \sqrt{h_{\mathrm{s}}^2 - 10h_{\mathrm{s}} +1} \right)$. 
Here, $\alpha_{1}$, $\alpha_{2}$ and $\delta$ are constants. 
There also exists the following model: 
$
\mathcal{F}(R,G)=\frac{\alpha}{(\tilde{f}(R,G))^{x+2}}R+\frac{\delta}{(\tilde{f}(R,G))^{x}}G-\frac{6h_{\mathrm{s}}}{(\tilde{f}(R,G))^{x+4}}\left[4h_{\mathrm{s}}^{2}\delta x+\alpha(x+2+h_{\mathrm{s}})\right]
$, where 
$
\tilde{f}(R,G)=\left\{\frac{-\alpha(x+2)R\pm
\sqrt{\alpha^{2}(x+2)^{2}R^{2}+24h_{\mathrm{s}}
\left[4h_{\mathrm{s}}^{2}\delta x
+\alpha(x+2+h_{\mathrm{s}})\right]
(x+4)\delta x G}}{2\delta x G}\right\}^{1/2}
$. 
Here, $\alpha$ and $x$ are constants. 

\noindent
(b) $\beta>1$ [Type I]:\ 
$
\mathcal{F}(R,G)=-4h_{\mathrm{s}}^{2}\lambda f(R,G)R+\lambda
\left(f(R,G)\right)^{1+2\beta}G+24h_{\mathrm{s}}^{4}\lambda \left(f(R,G)\right)^{1-2\beta}
$
with 
$
\tilde{f}(R,G)=
\left[\frac{h_{\mathrm{s}}^{2}R+\sqrt{h_{\mathrm{s}}^{4}R^{2}+6h_{\mathrm{s}}^{4}(4\beta^{2}-1
)G}}{(1/2+\beta)G}\right]^{1/\left(2\beta\right)}
$, where $\lambda$ is a constant. 

\noindent
(c) $\beta<1$ [Type II ($-1<\beta<0$), Type III ($0 < \beta < 1$), and 
Type IV ($\beta<-1$ but $\beta$ is not an integer)]:\ 
$\mathcal{F}(R,G)=R+\left(3/2\right)\left(G/R\right)$. 
In Ref.~\cite{Bamba:2009uf}, it has been examine that 
the finite-time future singularities 
can be removed by the term 
$R^{\vartheta_1}G^{\vartheta_2}$, where $\vartheta_1 (>0)$ and $\vartheta_2 (>0)$ are positive integers.

\section{Dark energy from $F(R)$ gravity with the Lagrange multipliers}

In this section, we study $F(R)$ gravity with the Lagrange multiplier field. 
With $F_1(R)$ and $F_2 (R)$ arbitrary functions of $R$, 
the action is expressed as 
\begin{equation}
S = \int d^4 x \sqrt{-g} \left[ F_1(R) - \lambda_{\mathrm{L}}
\left( \frac{1}{2}
\partial_\mu R \partial^\mu R
+ F_2 (R) \right) \right]\,,
\label{eq:8}
\end{equation}
where $\lambda_{\mathrm{L}}$ is the Lagrange multiplier field and 
yields a constraint equation $\left( 1/2\right) \partial_\mu R \partial^\mu R + F_2 (R) = 0$. 
The variation of the action in Eq.~({\ref{eq:8}}) with respect to $g_{\mu\nu}$ 
leads to the gravitational field equation as 
\begin{eqnarray}
&& \hspace{-10mm}
\frac{1}{2} g_{\mu\nu} F_1(R) 
+ \frac{1}{2} \lambda_{\mathrm{L}} \partial_\mu R \partial_\nu R 
+ \left( -
R_{\mu\nu} + \nabla_\mu \nabla_\nu - g_{\mu\nu} \Box 
\right) 
\nonumber \\
&& \hspace{20mm}
{}\times 
\left[ \frac{d F_1(R)}{dR} - \lambda_{\mathrm{L}} \frac{d F_2(R)}{dR} - \nabla^\mu 
\left(\lambda_{\mathrm{L}} 
\nabla_\mu R \right) \right] =0\,,
\label{eq:9}
\end{eqnarray}
where 
${\nabla}_{\mu}$ is the covariant derivative 
and $\Box \equiv g^{\mu \nu} {\nabla}_{\mu} {\nabla}_{\nu}$
is the covariant d'Alembertian. 
For the de Sitter space-time, which 
realize the current cosmic accelerated expansion, i.e., 
the dark energy dominated stage, 
the scalar curvature is a positive constant 
value $R_0$ and hence the Ricci tensor becomes 
$R_{\mu\nu} = \left( 1/4\right) R_0 g_{\mu\nu}$. 
In this case, from the above constraint equation and Eq.~({\ref{eq:9}}), 
we have 
$
\lambda_{\mathrm{L}} = \left[ - 2 F_1 (R_0) + R_0
\left(d F_1(R_0)/dR\right) \right]/\left[ R_0 \left(d F_2(R_0)/dR\right) 
\right]
$. 
Moreover, 
in the flat FLRW background, the above constraint equation reads 
$-\left(1/2\right) \dot{R}^2 + F_2(R) =0$. 
For $F_2(R) >0$, this equation can be solved in terms of $t$ 
as $t = \int^R dR/\sqrt{2F_2(R)}$. 
Provided that the form of $H(t)$ is given by the analysis of 
the observational data, $F_2(R)$ is able to be reconstructed so that 
the evolution of $H(t)$ can be reproduced. 
It follows from $R=6\left[ dH/dt +2H^2 \right]$ that $H(t)$ presents 
the evolution of $R = R(t)$, and by solving this equation inversely, 
we can find $t=t(R)$. Thus, we acquire 
$F_2(R) = \left(1/2\right) 
\left(d R/dt\right)^2$ with $t=t(R)$. 
We note that $F_1(R)$ is an arbitrary function of $R$. 
As an example, we consider $H(t) = h_0/t$ with $h_0 >1$ 
leading to $a(t) = a_0 t^{h_0}$, where $h_0$ and $a_0$ are constants. 
In this case, the accelerated expansion of the universe or 
power-law inflation happens. 
We have $R = 6h_0 \left(-1 + 2h_0\right)/{t^2}$, from which we 
also acquire $t = \sqrt{6h_0 \left(-1 + 2h_0\right)/R}$. 
Using these relations, we obtain 
$F_2(R) = R^3/\left[12 h_0 \left( - 1 + 2 h_0 \right)\right]$. 
As another example, we examine the case that $R$ is described by 
$R = \left(R_-/2\right) \left( 1 - \tanh \omega t \right)
+ \left(R_+/2\right) \left( 1 + \tanh \omega t \right)$ 
with $R_{\pm} (>0)$ and $\omega (>0)$ positive constants. 
In the limit $t \to \pm \infty$, $R \to \pm R_{\pm}$, 
and therefore the universe asymptotically approaches the de Sitter space-time. 
In this case, we can regard that in the limit $t \to - \infty$, 
inflation in the early universe occurs, whereas that 
in the limit $t \to + \infty$, the late-time cosmic acceleration happens. 
We also have 
$
F_2(R) = \left(1/8\right) \left(R_- - R_+\right)^2 \omega^2 \left[ 1
  - \left( R_- + R_+ - 2R \right)^2/\left( R_- - R_+ \right)^2
\right]^2
$. As a consequence, for the above $R$, this $F(R)$ gravity model with 
the constraint originating from the Lagrange multiplier 
can be a unified scenario between inflation and dark energy era, 
although it should carefully be studied whether 
the reheating stage after inflation can be realized.  
Furthermore, $F_1(R)$ does not affect cosmological evolution of the 
universe and influences only the correction of 
the Newton law. Thus, cosmology is determined only by the form of $F_2(R)$. 

To explore the Newton law, 
we take $F_1(R) = R/\left(2\kappa^2\right)$ as the Einstein-Hilbert term 
and add matter. 
In this case, 
for $\lambda_{\mathrm{L}} = 0$, from Eq.~({\ref{eq:9}}) we find 
the Einstein equation 
$\left[
R_{\mu\nu} - \left(1/2\right) g_{\mu\nu} R \right] = \kappa^2 T^{(\mathrm{M})}_{\mu\nu}$ 
with $T^{(\mathrm{M})}_{\mu\nu}$ the energy-momentum tensor of matter 
Its trace equation reads $R=-\kappa^2 T^{(\mathrm{M})}$, where 
$T^{(\mathrm{M})}$ is the trace of $T^{(\mathrm{M})}_{\mu\nu}$. 
Moreover, the constraint equation is given by 
$\left(\kappa^4/2\right) \partial_\mu
T \partial^\mu T + F_2 \left( - \kappa^2 T \right)=0$. 
Since this is not always met, we should modify the constraint equation as 
$
\left(1/2\right) \partial_\mu R \partial^\mu R + F_2(R) 
- \left(\kappa^4/2\right)  \partial_\mu T \partial^\mu T^{(\mathrm{M})} - F_2
\left( - \kappa^2 T \right) = 0
$. Thus, this implies that the action with the constraint coming 
from the Lagrange multiplier field and matter should be described by 
\begin{eqnarray}
S \Eqn{=} \int d^4
x \sqrt{-g} \left\{ \frac{R}{2\kappa^2} - \lambda_{\mathrm{L}} \left[ 
\frac{1}{2} \partial_\mu R \partial^\mu R + F_2 (R) -
\frac{\kappa^4}{2} \partial_\mu T^{(\mathrm{M})} \partial^\mu T^{(\mathrm{M})} 
\right. 
\right. 
\nonumber \\ 
&& \left. \left. 
{}- F_2 \left( -\kappa^2 T^{(\mathrm{M})} \right) \right] 
+ \mathcal{L}_{\mathrm{M}}
\right\}\,,
\label{eq:10}
\end{eqnarray}
For the case of the vacuum such that $T^{(\mathrm{M})} =0$, 
the constraint equation is 
$ 
\left(1/2\right) \partial_\mu R \partial^\mu R + F_2
(R) - F_2 \left( 0 \right) =0
$. 
If $F_2 \left( 0 \right)=0$, e.g., the first example 
of $F_2(R) = R^3/\left[12 h_0 \left( - 1 + 2 h_0 \right)\right]$ 
shown above, 
this is equivalent to the 
constraint equation derived from the action in Eq.~({\ref{eq:8}}).  
In this case, there exist two types of the solutions in 
the constraint equation $-\left(1/2\right) \dot{R}^2 + F_2(R) =0$. 
One is $R=0$ and the other is presented by 
$t = \int^R dR/\sqrt{2F_2(R)}$. 
On the small scales of, e.g., the solar system and galaxies, the 
solution would be the first solution of $R=0$ so 
that the Newton law can be recovered. 
On the other hand, in the bulk of the universe, 
the solution should be $t = \int^R dR/\sqrt{2F_2(R)}$ in order 
that the cosmic evolution can be realized. 
It is not so clear 
whether the first solution 
on the small scales of the solar system and galaxies and 
the second one in the bulk universe can be 
connected in the intermediate scales.

\vspace{-2mm}

\section{Domain wall solutions in $F(R)$ gravity}

In this section, we investigate a static domain wall solution 
and reconstruct an $F(R)$ gravity model with realizing it~\cite{Bamba:2011nm}. 

\vspace{-0.5mm}

\subsection{Static domain wall solution in a scalar field theory}

To begin with, we study a static domain wall solution in a scalar field 
theory. 
We suppose that 
the following $D=d+1$ dimensional warped metric
$
ds^2 = dy^2 + \e^{u(y)} \sum_{\mu,\nu=0}^{d-1} {\hat g}_{\mu\nu} 
dx^\mu dx^\nu
$, 
and that the scalar field only depends on $y$. 
In this background, 
the metric in the $d$-dimensional Einstein manifold 
is ${\hat g}_{\mu\nu}$, defined by 
${\hat R}_{\mu\nu} = \left[ \left(d-1\right)/l^2 \right] {\hat g}_{\mu\nu}$. 
In addition, for $1/l^2>0$, the space is the de Sitter one, 
for $1/l^2>0$, it is the anti-de Sitter, 
and for $1/l^2 = 0$, it is the flat. 
Following the procedure proposed in Ref.~\cite{Capozziello:2005tf}, 
it has been demonstrated that 
a static domain wall solution can exist in a scalar field theory~\cite{Bamba:2011nm} (a developed study on a static domain wall solution in a scalar field theory has also been executed in Ref.~\cite{Toyozato:2012zh}). 
We investigate the action 
$S = \int d^D x \sqrt{-g} \left[ \left(R/2\kappa^2\right) - 
\left(1/2\right) \omega(\varphi) \partial_\mu \varphi \partial^\mu \varphi 
 - \mathcal{V} (\varphi) \right]$, where $\omega(\varphi)$ is 
a function of the kinetic term of a scalar field $\varphi$ and 
$\mathcal{V} (\varphi)$ is a potential of $\varphi$. 
In the above $D$ dimensional warped metric, with 
the $(y,y)$ and $(\mu,\nu)$ components of the Einstein equation, 
we obtain the expressions of $\omega(\varphi)$ and $\mathcal{V} (\varphi)$. 
Using these expressions, the energy density is described as 
$
\rho_{\varphi} \equiv \left(1/2\right) \omega(\varphi) \left( \varphi' \right)^2 + \mathcal{V}(\varphi)
$. 
As an example, we consider 
$u = u_0 \exp \left(-y^2/y_0^2\right)$, 
where $u_0$ and $y_0$ are constants. 
In this case, 
the distribution of $\rho_{\varphi}$ reads 
$
\rho_{\varphi} (y) = -\left[\left(d-1\right)/\left(2y_0^2\right)\right] 
\left[\left( 2y^2/y_0^2 \right) - 1 \right] \exp\left(- y^2/y_0^2 \right) 
+ \left[\left(d-1\right)^2/l^2\right] \exp\left[ -u_0 \exp\left(-y^2/y_0^2 
\right)\right]
$. 
Accordingly, the energy density of $\varphi$ is localized at $y\sim 0$ and 
thus a domain wall is made. We note that a condition for $\rho_{\varphi}$ 
to be localized is $u \to 0$ in the limit $\left| y \right| \to \infty$.

\subsection{Reconstruction of the form of $F(R)$}

In the $D$ dimensional warped metric, 
the $(y,y)$ component and the trace of $(\mu,\nu)$ components of 
the gravitational field equation read 
%
\begin{eqnarray} 
&&
\hspace{-10mm}
\frac{d-1}{2} u' \left( F_{,R} \right)' 
-\frac{d}{2} \left[ u'' + \frac{1}{2}\left(u'\right)^2 
\right] F_{,R} -\frac{1}{2} F = \kappa^2 T^{(\mathrm{M})}_{yy}\,, 
\label{eq:11} \\ 
%
&&
\hspace{-10mm}
d \left( F_{,R} \right)'' 
+ \frac{d\left(d-2\right)}{2} u' 
\left( F_{,R} \right)' 
+ \left\{ 
-\frac{d}{2} \left[ u'' 
+ \frac{d}{2}\left(u'\right)^2 \right] 
+ \frac{d\left(d-1\right)}{l^2} \e^{-u}
\right\} F_{,R} 
\nonumber \\ 
&&
\hspace{-10mm}
{}-\frac{d}{2} F 
= \kappa^2 
\sum_{\mu,\nu=0}^{d-1} 
g^{\mu\nu} T^{(\mathrm{M})}_{\mu\nu}\,,
\label{eq:12}
\end{eqnarray}
where the prime denotes the derivative with respect to $y$ of $d/dy$, 
and $\left( F_{,R} \right)' \equiv d F_{,R}/d y$ 
and $\left( F_{,R} \right)'' \equiv d^2 F_{,R}/d y^2$. 
We examine an explicit form of $F(R)$ with leading to a domain wall solution 
for the case that matter is absent. 
For the model 
$
u = u_0 \exp\left(-y^2/y_0^2\right)
$, 
with the relation 
$
R = -d \left\{ u'' 
+ \left[\left(1+d\right)/4\right] \left(u'\right)^2 \right\} 
+ \left[d\left(d-1\right)/l^2\right] \e^{-u}
$, 
$y$ can be described as a function of $R$, $y=y(R)$, 
and eventually we find $u=u(y(R))$. 
By plugging this equation into Eqs.~(\ref{eq:11}) and (\ref{eq:12}) 
and eliminating $y$, 
Eqs.~(\ref{eq:11}) and (\ref{eq:12}) can be expressed 
as differential equations in terms of $F(R)$. 
Here, it is enough to analyze Eq.~(\ref{eq:11}) because 
Eq.~(\ref{eq:12}) is not independent of Eq.~(\ref{eq:11}). 
As a result, Eq.~(\ref{eq:11}) can be rewritten to 
\begin{eqnarray} 
&&
{\Xi}_1 (R) \frac{d^2 F(R)}{d R^2} + {\Xi}_2 (R) \frac{d F(R)}{d R} 
- F(R) = 0\,, 
\label{eq:13} \\
&&
{\Xi}_1 (R) \equiv \left( d-1 \right) u' \frac{d R}{d y} 
= \left( d-1 \right) \left( \frac{d R}{d y} \right)^2 
\frac{d u(y(R))}{d R}\,, 
\label{eq:14} \\ 
&&
{\Xi}_2 (R) \equiv 
\left( - d \right) \left[ u'' + \frac{1}{2}\left(u'\right)^2 \right] 
= \left( - d \right) 
\left[  
\frac{d^2 R}{d y^2} \frac{d u(y(R))}{d R} 
\right. 
\nonumber \\ 
&& \left. 
\hspace{10mm}
{}+ \left( \frac{d R}{d y} \right)^2 \frac{d^2 u(y(R))}{d R^2} 
+ \frac{1}{2} \left( \frac{d R}{d y} \right)^2 
\left( \frac{d u(y(R))}{d R} \right)^2 \right]\,. 
\label{eq:15} 
\end{eqnarray}
To solve the above relation of $R$ in terms of $y$, 
by defining $Y \equiv y^2/y_0^2$ and 
expanding exponential terms in the limit $Y = y^2/y_0^2 \ll 1$, 
we take only the first leading terms in terms of $Y$. 
We find 
$
Y = y^2/y_0^2 \approx \left(R - \gamma_1\right)/\gamma_2
$ with 
$
\gamma_1 \equiv \left( 2d u_0/y_0^2 \right) + d\left(d-1\right)/l^2 
$ 
and 
$
\gamma_2 \equiv -d \left( u_0/y_0^2 \right) \left[ 6+\left(1+d\right) u_0 
\right] + \left[ d\left(d-1\right)/l^2 \right] u_0 
$, 
where $\gamma_1$ and $\gamma_2$ are constants. 
Finally, for $Y = y^2/y_0^2 \ll 1$, 
Eq.~(\ref{eq:13}) can be described by 
$
\left(d^2 F(R)/d R^2\right) + \mathcal{C} \left( d F(R)/d R\right) + 
\mathcal{D} F(R) = 0
$ 
with 
$ 
\mathcal{C} \equiv {\Xi}^{(0)}_2 / {\Xi}^{(0)}_1 
$ and 
$
\mathcal{D} \equiv -1/{\Xi}^{(0)}_1
$, 
where ${\Xi}^{(0)}_1$ and ${\Xi}^{(0)}_2$ constants described by 
the model parameters $d$, $l$, $u_0$ and $y_0$. 
We acquire a general solution of this equation as 
$
F(R) = F_{+} \e^{\lambda_{+} R} + F_{-} \e^{\lambda_{-} R}
$, 
where 
$
\lambda_{\pm} \equiv \left(1/2\right) 
\left( -\mathcal{C} \pm \sqrt{{\mathcal{C}}^2-4\mathcal{D}} \right) 
$, and $F_{\pm}$ are arbitrary constants. 
Here, the subscriptions $\pm$ of $\lambda_{\pm}$ correspond to the sign 
``$\pm$'' 
on the right-hand side 
of this equation. 
In 
the model 
$
u = u_0 \exp\left(-y^2/y_0^2\right)
$, at $y \sim 0$ 
the distribution of the energy density is localized and therefore 
a domain wall is realized as shown above. 
Consequently, for an exponential model of $F(R)$ gravity, 
a domain wall can appear at $y \sim 0$.

\subsection{Effective (gravitational) domain wall} 

Next, with the reconstruction method~\cite{Reconstruction, Nojiri:2006gh}, 
we explore an effective (gravitational) domain wall in $F(R)$ gravity.
With the same procedure as in Section 2.3,   
we study the action of $F(R)$ gravity given by $\mathcal{F}(R,G) = F(R)$. 
Using two proper functions $P(\psi)$ and $Q(\psi)$ of a scalar field 
$\psi$, we represent the term $\mathcal{F}(R,G) = P(\psi)R+Q(\psi)$. 
The variation over $\psi$ yields 
$\left( d P(\psi)/d \psi \right) R + d Q(\psi)/d \psi = 0$. 
Solving this equation with respect to $\psi$ leads to $\psi=\psi(R)$, 
by substituting which into the action in Eq.~(\ref{eq:1}) 
with $\mathcal{F}(R,G) = P(\psi)R+Q(\psi)$, we acquire the 
action of $F(R)$ gravity as $F(R) = P(\psi(R)) R + Q(\psi(R))$. 
In the $D$ dimensional warped metric shown in Section 4, 
for the case that $\psi$ depends only on $y$, 
it follows from the gravitational field equation with the choice of 
$\psi=y$ and $1/l^2 =0$ (i.e., the flat space), we have 
\begin{eqnarray}
&&
u'(\psi) = - \frac{2}{d-1} 
\left[ 
\frac{P'(\psi)}{P(\psi)} + \frac{d}{d-1} 
\left(P(\psi)\right)^{1/\left(d-1\right)} 
\right. 
\nonumber \\ 
&& \left. 
\hspace{28mm}
{}\times 
\int d\psi \left(P(\psi)\right)^{-\left(2d-1\right)/\left(d-1\right)} \left(P'(\psi)\right)^2
\right],
\label{eq:16} \\
&&
Q(\psi) =
\frac{d(d-1) \left(u'(\psi)\right)^2}{4} P(\psi) + (d-1) u'(\psi) P'(\psi)\,, 
\label{eq:17}
\end{eqnarray}
where the prime denotes the derivative with respect to $\psi (= y)$ of 
$d/d \psi$. 
For a model $P(\psi) = \left(U(\psi)\right)^{-2(d-1)}$ and 
$U(\psi) = U_0 \left( \psi^2 + \psi_0^2 \right)^\chi$ with 
$U_0$, $\psi_0$ and $\chi$ constants, we acquire 
%
\begin{equation}
u'(\psi) = \frac{2\chi \psi}{\psi^2 + \psi_0^2} 
 - \frac{32 d \chi^2 \psi^{4\chi -1} }{\left( \psi^2 + \psi_0^2 \right)^{2\chi}}\nonumber 
\end{equation}
\begin{equation}
\times 
\sum_{k=0}^\infty 
\frac{\Gamma\left(2\chi - 1 \right)}{\left(4\chi -1 - 2k\right) \Gamma\left( 2\chi - 1 - k \right) k!}
\left(\frac{\psi_0^2}{\psi^2}\right)^k\,.
\label{eq:18}
\end{equation}
In the range where $\psi=y$ is large, 
we take $\chi= -1/\left[4\left(4d -1\right)\right]$ 
and impose the boundary condition that in the limit $|y|=|\psi| \to \infty$, 
the universe asymptotically approaches flat as $u \to 0$. 
As a result, we obtain 
$
u(\psi)  = - \left\{1/\left[4 \left(6 d -1\right)\right]\right\} 
\left(\psi_0/\psi\right)
+ \mathcal{O} \left( \left(\psi_0/\psi\right)^3\right) 
$. {}From this expression, we see that 
$u(\psi)$ performs a non-trivial behavior at $\psi = y \sim 0$. 
Hence, it can be considered that an effective 
(gravitational) domain wall could appear at $y=0$. 
Moreover, by using the representation 
$
u'(\psi) = \left(4U'(\psi)/U(\psi)\right) - \left(8d/U(\psi)^2 \right)
\int d\psi \left(U'(\psi)\right)^2
$, 
we acquire an integration expression of $u(\psi)$ as 
\begin{eqnarray} 
&&
\hspace{-10mm}
u(\psi) = 8 \chi \int_{-\infty}^{\psi} d \psi \frac{\psi}{\psi^2 + \psi_0^2} 
\nonumber \\
&& \hspace{5mm}
{}-32 d \chi^2 \int_{-\infty}^{\psi} d \psi \frac{1}{\left( \psi^2 + \psi_0^2 
\right)^{2 \chi}} \int_{0}^{\psi} d \tilde{\psi} 
\left( \tilde{\psi}^2 + \psi_0^2 \right)^{2 \left( \chi - 1 \right)} 
\tilde{\psi}^2\,.
\label{eq:19}
\end{eqnarray}
In Ref.~\cite{Bamba:2011nm}, it has numerically been verified that 
there exists a local maximum of $u(\psi)$ 
at $\psi = y \sim 0$, 
and thus an effective 
(gravitational) domain wall could be realized at $y=0$. 
Also, there occurs such a qualitative behavior of $u(\psi)$ in terms of $\psi$ 
regardless of the values of the model parameters. 
In addition, we mention that 
for $U(\psi) = U_0 \sqrt{ \psi^2 + \psi_0^2}$, i.e., 
$\chi = 1/2$, there exists an analytic solution 
\begin{equation} 
u(\psi) = 2\left( 1-2d \right) \ln \left( \psi^2 + \psi_0^2 \right) +4d 
\left( \arctan \left( \frac{\psi}{\sqrt{\psi_0^2}} \right) \right)^2 + 
\mathcal{C}\,, 
\label{eq:20}
\end{equation}
with $\mathcal{C}$ an integration constant. 
In this case, 
for the region of a small amplitude of $\psi$, 
it is considered that the distribution of the energy density 
is localized, so that 
an effective (gravitational) domain wall could be made.  

As a demonstration, 
we reconstruct an explicit $F(R)$ form 
for $u(\psi)$ in Eq.~(\ref{eq:20}), although 
only in the region of a small amplitude of $\psi$, 
the distribution of the energy density could be regarded as  
an effective (gravitational) domain wall. 
{}From $P'(\psi)R + Q'(\psi) = 0$ 
and Eq.~(\ref{eq:17}), we have 
\begin{equation}  
R = -\frac{Q'(\psi)}{P'(\psi)} = - \frac{\left( d-1 \right)}{2 P'(\psi)} 
\left( d u'(\psi) u''(\psi) + 2u''(\psi)P'(\psi)  + 2u'(\psi) P''(\psi)
\right)\,. 
\label{eq:21}
\end{equation}
Solving this equation, an analytic relation 
$\psi = \psi(R)$ can be found. With this relation, we acquire 
$F(R) = P(\psi(R)) R + Q(\psi(R))$. 
In this case, we have 
$P(\psi) = \left( U_0 \psi_0 \right)^{-2\left( d-1 \right)} 
\left( 1+\bar{Y} \right)^{-2\left( d-1 \right)}$. 
For $\bar{Y} \equiv \psi^2/ \psi_0^2 \ll 1$, 
by expanding Eq.~(\ref{eq:17}) in terms of $\bar{Y}$ 
and taking the leading terms, we find 
$R = {\mathcal{R}}_0 + {\mathcal{R}}_1 \bar{Y}$. 
Furthermore, from Eq.~(\ref{eq:17}) we also obtain
$Q = {\mathcal{Q}}_1 \bar{Y} + {\mathcal{Q}}_2 \bar{Y}^2$. 
Here, ${\mathcal{R}}_0$, ${\mathcal{R}}_1$, 
${\mathcal{Q}}_1$ and ${\mathcal{Q}}_2$ are constants and 
these are written by using the model parameters 
$d$, $U_0$ and $\psi_0$. 
Moreover, with $R = {\mathcal{R}}_0 + {\mathcal{R}}_1 \bar{Y}$ 
we describe 
$\bar{Y} = \bar{Y}_0 + \bar{Y}_1 R$, where 
$\bar{Y}_0 \equiv -{\mathcal{R}}_0/{\mathcal{R}}_1$ and 
$\bar{Y}_1 \equiv 1/{\mathcal{R}}_1$. 
$P(\psi)$ can also be expanded as 
$P(\psi) \approx \left( U_0 \psi_0 \right)^{-2\left( d-1 \right)} 
\left\{ 1-\left( d-1 \right)\bar{Y} + 
\left[d\left( d-1 \right)/2\right]\bar{Y}^2 \right\}$. 
By plugging this relation and 
$Q = {\mathcal{Q}}_1 \bar{Y} + {\mathcal{Q}}_2 \bar{Y}^2$
with $\bar{Y} = \bar{Y}_0 + \bar{Y}_1 R$ 
into $F(R) = P(\psi(R)) R + Q(\psi(R))$ 
and taking terms of order of $R^2$, 
we find $F(R) = {\mathcal{F}}_0 + {\mathcal{F}}_1 R + {\mathcal{F}}_2 R^2$, 
where ${\mathcal{F}}_0$, ${\mathcal{F}}_1$ and ${\mathcal{F}}_2$ are 
constants and these are represented by 
the model parameters $d$, $U_0$ and $\psi_0$. 
The above explicit form of $F(R)$ has been derived 
for $\bar{Y} = \psi^2/ \psi_0^2 \ll 1$. 
Hence, it follows from $R = {\mathcal{R}}_0 + {\mathcal{R}}_1 \bar{Y}$ 
that this $F(R)$ form can be considered to correspond to the one 
for $R \sim \mathcal{O} (1)$ if ${\mathcal{R}}_0 \sim \mathcal{O} (1)$. 
Therefore, when we choose ${\mathcal{F}}_0 = 0$ and ${\mathcal{F}}_1 = 1$, 
we have $F(R) = R + {\mathcal{F}}_2 R^2$. 
In this case, 
for the small curvature limit, 
$F(R)$ approaches $R$, i.e., general relativity, asymptotically. 
As a result, if $u(\psi)$ is given by Eq.~(\ref{eq:20}) in which an effective 
(gravitational) domain wall can be realized, 
an explicit form of $F(R)$ is expressed as a power-law model. 
We state the difference between the domain walls in Sections 4.2 and 4.3. 
A pure gravitational effect yields 
an effective (gravitational) domain wall in Section 4.3, 
but a scalar field makes 
a static domain wall solution explored in Section 4.1. 
In Section 4.2, the deviation of $F(R)$ gravity from general 
relativity is equivalent to matter geometrically, i.e., 
a scalar field in Section 4.1.

\section{Covariant higher derivative gravity with scalar projectors} 

It is considered that a covariant gravity which is 
power-counting renormalizable would be 
higher derivative theory, e.g., models in Ref.~\cite{C-R-H-G}. 
Higher derivative gravity is very well known to be renormalizable 
multiplicatively (for a review, see, for example,~\cite{Buchbinder:1992rb}). 
However, in general, such a higher derivative theory cannot 
keep the unitarity. To retain it, 
the so-called Ho\v{r}ava gravity~\cite{Horava:2009uw} 
has been proposed. 
In this section, we make the formulation for covariant 
higher derivative gravity 
with Lagrange multiplier constraint as well as scalar projectors. 
In particular, we construct a gravity theory with the Lorentz symmetry and/or
the full general covariance in the action, 
although these symmetry and/or covariance is spontaneously broken. 
In such a theory, the propagator of the graviton in the 
ultraviolet (UV) region can be improved better, whereas 
there appears no extra mode such as a scalar one.

\subsection{Model}

We explore the following action with the Lagrange multiplier field $\lambda_{\mathrm{L}}$~\cite{Capozziello:2010uv, L-M-F} 
as well as the scalar field $\Phi$: 
$
S_{\mathrm{L}} = - \int d^4 x \sqrt{-g} \lambda_{\mathrm{L}} \left[ \left(1/2\right) \partial_\mu \Phi \partial^\mu \Phi + \bar{W} \right]
$, where $\mu$ and $\nu$ run $0, 1, 2, 3$ and the $0$ component denotes 
the time $t$ as $\partial_0 \equiv \partial/\partial t$. {}From this action, 
we find a constraint equation 
$\left(1/2\right) \partial_\mu \Phi \partial^\mu \Phi
+ \bar{W} = 0$. This means that the vector quantity 
$\left( \partial_\mu \Phi \right)$ is time-like one. 
Hence, this breaks
the Lorentz symmetry and/or the full general covariance 
spontaneously. 
For simplicity, we assume that $\bar{W}$ is a constant, although 
this assumption is not necessary for the symmetry and/or covariance to 
spontaneously be broken. 
Furthermore, the direction of time can be taken so that it should be 
parallel to the vector quantity $\left( \partial_\mu \Phi \right)$. 
In this case, we have 
$\left(1/2\right) \left( d \Phi/d t \right)^2 = \bar{W}$, from which we have 
$\Phi = \sqrt{2\bar{W}}t$. Accordingly, the spatial region is a hypersurface 
with a constant $\Phi$ because the hypersurface becomes orthogonal to 
the vector quantity $\left( \partial_\mu \Phi \right)$. 
In the flat space-time, we examine the perturbations as 
$g_{\mu\nu} = \eta_{\mu\nu} + h_{\mu\nu}$ with $\eta_{\mu\nu}$ the Minkowski 
metric and $h_{\mu\nu}$ corresponds to the fluctuations, i.e., the deviation 
of $g_{\mu\nu}$ from the Minkowski background $\eta_{\mu\nu}$. 
A projection operator is defined as 
$
\mathcal{P}_\mu^{\ \nu} \equiv \delta_\mu^{\ \nu} + 
\left(\partial_\mu \Phi \partial^\nu \Phi \right)/\left(2\bar{W} \right)
$ with $\mathcal{P}_0^{\ \mu} = 0$. 
As a result, 
the action describing a higher derivative gravity with scalar projector 
which is covariant and power-counting renormalizable is expressed as 
\begin{eqnarray}
\hspace{-7.5mm}
&&
S_{2n+2} =
\int d^4 x \sqrt{-g} \left\{ \frac{R}{2\kappa^2} 
- \zeta 
\left[
\left(\partial^\mu \Phi \partial^\nu \Phi \nabla_\mu
\nabla_\nu - \partial_\mu \Phi \partial^\mu \Phi
\nabla^\rho \nabla_\rho \right)^n
\mathcal{P}_\alpha^{\ \mu} \mathcal{P}_\beta^{\ \nu} 
\right. \right. 
\nonumber \\
\hspace{-7.5mm}
&& \left. \left.
\times 
\left( R_{\mu\nu} - \frac{1}{2 \bar{W}}
\partial_\rho \Phi \nabla^\rho \nabla_\mu \nabla_\nu \Phi
\right) \right] 
\left[
\left(\partial^\mu \Phi \partial^\nu \Phi \nabla_\mu
\nabla_\nu - \partial_\mu \Phi \partial^\mu \Phi
\nabla^\rho \nabla_\rho \right)^n
\right. \right. 
\nonumber \\
\hspace{-7.5mm}
&& \left. \left. 
\times
\mathcal{P}^{\alpha\mu} \mathcal{P}^{\beta\nu} 
\left( R_{\mu\nu} 
- \frac{1}{2 \bar{W}}
\partial_\rho \Phi \nabla^\rho \nabla_\mu \nabla_\nu \Phi
\right) \right] 
- \lambda_{\mathrm{L}} \left( \frac{1}{2} \partial_\mu \Phi \partial^\mu \Phi
+ \bar{W} \right) \right\}\,, 
\label{eq:22} 
\\ 
%
\hspace{-7.5mm}
&&
S_{2n+3} = \int d^4 x \sqrt{-g} \left\{ \frac{R}{2\kappa^2} - \zeta
\left[
\left(\partial^\mu \Phi \partial^\nu \Phi \nabla_\mu
\nabla_\nu - \partial_\mu \Phi \partial^\mu \Phi
\nabla^\rho \nabla_\rho \right)^n
\mathcal{P}_\alpha^{\ \mu} \mathcal{P}_\beta^{\ \nu} 
\right. \right. 
\nonumber \\
\hspace{-7.5mm}
&& \left. \left.
\times 
\left( R_{\mu\nu} - \frac{1}{2 \bar{W}}
\partial_\rho \Phi \nabla^\rho \nabla_\mu \nabla_\nu \Phi
\right) \right] 
\left[
\left(\partial^\mu \Phi \partial^\nu \Phi \nabla_\mu
\nabla_\nu - \partial_\mu \Phi \partial^\mu \Phi
\nabla^\rho \nabla_\rho \right)^{n+1}
\right. \right. 
\nonumber \\
\hspace{-7.5mm}
&& \left. \left. 
\times 
\mathcal{P}^{\alpha\mu} \mathcal{P}^{\beta\nu} \left( R_{\mu\nu} - \frac{1}{2 \bar{W}}
\partial_\rho \Phi \nabla^\rho \nabla_\mu \nabla_\nu \Phi
\right) \right] 
- \lambda_{\mathrm{L}} \left( \frac{1}{2} \partial_\mu \Phi \partial^\mu \Phi
+ \bar{W} \right) \right\}\,.
\label{eq:23}
\end{eqnarray}
Here, Eqs.~(\ref{eq:22}) and (\ref{eq:23}) are 
 for $z = 2n+2$ and $z = 2n+3$ ($n=0, 1, 2, \dotsc$), 
respectively, where $z$ is the quantity denoting 
the anisotropy between the time and spatial coordinates~\cite{Horava:2009uw}. 
Moreover, 
the gravitational field equation is given by 
$
\left[1/\left(2\kappa^2\right)\right] 
\left[ R_{\mu\nu} - \left(1/2\right) g_{\mu\nu} R \right]
+ G^{(\mathrm{higher})}_{\mu\nu} - \left(\lambda_{\mathrm{L}}/2\right) 
\partial_\mu \Phi \partial_\nu \Phi + \left(1/2\right) 
g_{\mu\nu}
\left[ \left(1/2\right) \partial_\rho \Phi \partial^\rho \Phi + \bar{W} \right]
=0 
$
with 
$G^{(\mathrm{higher})}_{\mu\nu}$ the higher derivative term, i.e., 
the second term, in the actions in Eq.~(\ref{eq:22}) and (\ref{eq:23}). 
Suppose the flat vacuum solution, 
the constraint equation 
$\left(1/2\right) \partial_\mu \Phi \partial^\mu \Phi
+ \bar{W} = 0$ becomes 
$\left(1/2\right) \left( d \Phi/d t \right)^2 = \bar{W}$. For the flat space solution, the gravitational field equation is reduced to 
$\lambda_{\mathrm{L}} \partial_\mu \Phi \partial_\nu \Phi =0$, 
because all of the term $\nabla_\mu
\nabla_\nu \Phi$ as well as the curvature terms vanish. 
The solution is given by $\lambda_{\mathrm{L}}=0$, because 
$\partial_\mu \Phi \neq 0$ owing to 
$\left(1/2\right) \partial_\mu \Phi \partial^\mu \Phi
+ \bar{W} = 0$. 
Thus, 
in these actions in Eqs.~(\ref{eq:22}) and (\ref{eq:23}) 
solutions with $\lambda_{\mathrm{L}}=0$ in the flat space vacuum 
can be realized. 
We further analyze the perturbations $h_{\mu\nu}$ with 
$\lambda_{\mathrm{L}}=0$. 
Using the diffeomorphism invariance in terms of the time coordinate, 
as a gauge condition for the unitarity, 
we set $\Phi = \sqrt{2\bar{W}}t$. 
By taking only the quadratic terms of the perturbations, 
we rewrite the actions in Eqs.~(\ref{eq:22}) and (\ref{eq:23}). 
At this stage, there remains 
the diffeomorphism invariance in terms of the spatial coordinates. 
In addition, the term $h_{0i}$ in the higher 
derivative term with a coefficient $\zeta$ does not exist 
in the rewritten actions. 
Moreover, the above constraint equation leads to $h_{00}=0$.  
We derive the equations by varying the actions in Eqs.~(\ref{eq:22}) and (\ref{eq:23}) with respect to $h_{00}$ and $\Psi$. 
We decompose $h_{0i}$ as $h_{0i} \partial_i s + v_i$ 
with $\partial^i v_i = 0$, 
where $s$ is the spatial scalar quantity and $v_i$ is a vector field. 
The invariance in terms of the spatial coordinates 
under the transformations of the linearized diffeomorphism is 
described as $\delta x^i = \partial^i u + w^i$ with 
$\partial_i w^i = 0$, 
where $u$ is the spatial scalar quantity and $w_i$ is a vector field. 
We find the following transformations under the diffeomorphism: 
$\delta s = \partial_t u$ and $\delta v_i = \partial_t w_i$. 
Accordingly, the gauge condition $s=v^i=0$, i.e., 
$h_{ti} = 0$ can be chosen. 
Furthermore, we express $h_{ij}$ as 
$
h_{ij} = \delta_{ij} A + \partial_j B_i + \partial_i B_j + C_{ij} + \left[
\partial_i \partial_j - \left(1/3\right) \delta_{ij}
\partial_k \partial^k \right] E
$, where $A$ and $E$ are scalar quantities, $B_i$ is a vector field, 
and $C_{ij}$ is a tensor field. 
Here, $\partial^i B_i = 0$, 
$\partial^i C_{ij} = \partial^j C_{ij} = 0$, and 
$C_i^{\ i}=0$. 
As a result, we acquire $\lambda_{\mathrm{L}} =0$ and 
vector $B_i = 0$, 
and thus the scalar $\lambda_{\mathrm{L}}$ and vector $B_i$ modes 
do not propagate. 
We fix the gauge in the actions rewritten above. 
We vary these actions over $A$ and $E$ and obtain equations. 
{}From these equations and $A = \left(1/3\right) \partial_k \partial^k E$, 
which is derived by the equation derived by the variation of the action with 
respect to $h_{0i}$ with the above decomposed representation of $h_{ij}$, 
we find $\partial_0^2 A =0$. 
Hence, since 
$A$ and $E$ have only the dependence on the spatial coordinate, 
these scalar quantities do not propagate. 
As a consequence, 
all the scalar modes $\Phi$, 
$\lambda_{\mathrm{L}}$, $h_{00}$, $s$, $A$ and $E$ 
and all the vector modes $v_i$ and $B_i$ do not propagate, 
whereas 
the propagating mode is only the tensor mode $C_{ij}$, namely, 
a massless graviton. 
This is a different feature from the Ho\v{r}ava 
gravity~\cite{Horava:2009uw} without the Lorentz invariance. 
The final expressions of the actions in Eqs.~(\ref{eq:22}) and (\ref{eq:23}) 
are given by 
\begin{eqnarray}
&&
S_{2n+2} = \int d^4 x \left\{ \frac{1}{8\kappa^2} \left[
C_{ij} \left( - \partial_0^2 + \partial_k \partial^k\right) C^{ij}
\right] 
\right. \nonumber \\ 
&& \left. 
{}- 2^{2n-2} \zeta \bar{W}^{2n} \left[\left( \partial_k \partial^k
\right)^{n+1}C_{ij}\right] 
\left[ \left( \partial_k \partial^k \right)^{n+1} C^{ij} \right] \right\}
\,, 
\label{eq:24} \\
%
%
&&
S_{2n+3} = \int d^4 x \left\{ \frac{1}{8\kappa^2} \left[
C_{ij} \left( - \partial_0^2 + \partial_k \partial^k\right) C^{ij}
\right] 
\right. \nonumber \\ 
&& \left. 
{}
- 2^{2n-1} \zeta \bar{W}^{2n+1} \left[ \left( \partial_k \partial^k
\right)^{n+1}C_{ij} \right] 
\left[ \left( \partial_k \partial^k \right)^{n+2} C^{ij}\right\} \right]
\,.
\label{eq:25}
\end{eqnarray}
Therefore, in the momentum space the propagator reads
\begin{eqnarray}
&& \left< h_{ij}(p) h_{kl}(-p) \right>
= \left< C_{ij}(p) C_{kl}(-p) \right> \nn
&& = \frac{1}{2} \left[ \left( \delta_{ij} - \frac{p_i
p_j}{\bm{p}^2}\right)
\left( \delta_{kl} - \frac{p_k p_l}{\bm{p}^2}
\right) - \left( \delta_{ik} - \frac{p_i p_k}{\bm{p}^2}\right)
\left( \delta_{jl} - \frac{p_j p_l}{\bm{p}^2}
\right) 
\right. \nonumber \\ 
&& \left. 
{}- \left( \delta_{il} - \frac{p_i p_l}{\bm{p}^2}\right)
\left( \delta_{jk} - \frac{p_j p_k}{\bm{p}^2}\right) \right] 
\nonumber \\ 
&& 
{}
\times \left\{
\begin{array}{ll}
1/\left[ p^2 - 2^{2n} \zeta \kappa^2 \bar{W}^{2n} \bm{p}^{4 \left( n+1 \right)} \right]
\, , & \mbox{for} \,\,\,\,\, z=2n+2\ \\
1/\left[ p^2 - 2^{2n-1} \zeta \kappa^2 \bar{W}^{2n+1} \bm{p}^{2 \left( 2n+3 \right)} \right]
\, , & \mbox{for} \,\,\,\,\, z=2n+3\ 
\end{array} \right. \, , 
\label{eq:26}
\end{eqnarray}
with ${\bm{p}}^2 = \sum_{i=1}^3 \left( p^i \right)^2$ and
$p^2 = - \left(p^0\right)^2 + {\bm{p}}^2$. 
If $\zeta>0$ and $p^0=0$, when 
$
2^{2n} \zeta \kappa^2 \bar{W}^{2n} \bm{p}^{4 n + 2} = 1
$ 
for $z=2n+2$ 
and 
$2^{2n-1} \zeta \kappa^2 \bar{W}^{2n+1} \bm{p}^{4 ( n+1 )} = 1$ 
for 
$z=2n+3$ 
are satisfied, 
the tachyonic pole exists. 
Thus, at least the flat vacuum is unstable. 
In this model, at least on the tree level, 
no propagating vector or scalar mode exists. 
The fact that the tensor structure of the propagator in 
Eq.~(\ref{eq:26}) changes 
implies that the vector or scalar mode could emerge. 
In other words, 
the vector or scalar mode has to be a composite state. 
At any perturbative level, this does not appear usually. 
Accordingly, the quantum corrections should not 
change the tensor structure. 
In the UV region with large $\bm{k}$, 
for $z=2$ ($n=0$) in Eq.~(\ref{eq:24}), 
the propagator evolves as $1/\left| \bm{k} \right|^4$.  
Hence, the UV behavior performs. 
While, 
for $z=3$ ($n=0$) in Eq.~(\ref{eq:25}), 
the propagator evolves as $1/\left| \bm{k} \right|^6$. 
Thus, the model is power-counting renormalizable.
For $z=2n +2$ ($n \geqslant 1$) in Eq.~(\ref{eq:24}) 
or $z=2n+3$ ($n \geqslant 1$) in Eq.~(\ref{eq:25}), 
the model is power-counting super-renormalizable. 
In the high energy regime, the dispersion relation of the graviton 
becomes $\omega = \bar{c} k^z$ with $\bar{c} (>0)$ a positive constant 
for the consistency of the dispersion relation. 
Here, 
$\omega$ is the angular frequency which corresponds to 
the energy and $k$ is the wave number which does to the momentum. 
Accelerating cosmology in such a theory was studied in Ref.~\cite{Kluson:2011rs}.

\section{Conclusions}

We have studied the accelerating (dark energy) solutions of modified gravity. 
These solutions may yield future singularities. 
We have explored the finite-time future singularities in 
$F(R)$, $F(G)$ and $\mathcal{F}(R,G)$ gravity theories. 
The removal of the finite-time future singularities in $F(R)$ gravity 
by adding an $R^2$-term which simultaneously leads to the 
unification of early-time inflation with late-time 
acceleration~\cite{Nojiri:2003ft} has been mentioned. 
The corresponding term may be different for $F(G)$ or $\mathcal{F}(R,G)$ 
gravity theory~\cite{Bamba:2009uf}. 
Moreover, we have studied dark energy in $F(R)$ gravity 
with the Lagrange multiplier field. 
Furthermore, domain wall solutions in $F(R)$ gravity have 
been presented. 
In addition, we have investigated 
covariant higher derivative gravity with scalar projectors.

\end{document}